\documentclass[aps,prb,reprint,amsmath,amssymb,floatfix]{revtex4-2}
\usepackage{graphicx}
\usepackage[colorlinks, citecolor=blue, linkcolor=blue]{hyperref}
\usepackage{txfonts}

\begin{document}

\title{Magnonic crystalline properties of stripe textures in thin ferromagnetic films}

\author{Joo-Von Kim}
\email{joo-von.kim@universite-paris-saclay.fr}
\author{Victor Leroy}
\author{Titiksha Srivastava}
\affiliation{Centre de Nanosciences et de Nanotechnologies, CNRS, Universit{\'e} Paris-Saclay, 91120 Palaiseau, France}
\date{\today}

\begin{abstract}
Ordered stripe domains in ferromagnetic thin films form a natural one‑dimensional crystal for propagating spin waves. Their spatial periodicity can be tuned readily with an applied magnetic field, making these systems an attractive platform for exploring how magnon band structures evolve as a function of the lattice constant $a$, a tuning that is difficult to achieve in physically patterned materials. In this work we employ micromagnetic simulations to calculate the spin‑wave spectra of a model iron‑garnet film, focusing on the influence of the external field and of cubic anisotropy. We find that band gaps at the Brillouin‑zone center ($k = 0$) and at the zone boundary ($k = \pm \pi/a$) respond differently to the applied field, appearing over a broad range of frequencies and wave vectors. When a perpendicular magnetic field or cubic anisotropy is present, additional gaps can appear at the middle of the reduced Brillouin zone ($k = \pm \pi/2a$). This behavior is interpreted as a Peierls‑type distortion of the domain‑wall lattice, wherein ``up'' and ``down'' domains alternately expand and contract under the influence of the effective perpendicular field.
\end{abstract}

\maketitle

\section{Introduction}
Spatially modulated phases develop in the order parameter as a result of competing short- and long-range interactions. In ferromagnetic thin films, the energy of competing exchange and dipolar interactions is minimized by organizing the magnetization into stripe domains, comprising of an ordered array of oppositely oriented domains separated by domain walls~\cite{Spreen:1978me}. The finer structure of such spin textures is determined by other energies, such as magnetocrystalline anisotropy, Zeeman, and Dzyaloshinskii-Moriya interactions. The interplay between the ensemble of these terms can result in parallel stripes, labyrinthine domains, and magnetic bubble lattices~\cite{HubertSchafer:1998md}.

Besides their intrinsic interest, spatially modulated phases have also found interest within magnonics~\cite{Yu:2021mt}, since ordered arrays of textures like domain walls or bubbles act like an artificial crystal for spin wave propagation. Drawing inspiration from advances in photonics and metamaterials, magnonic crystals have been studied as a means for designing magnon band structures for particular applications by tailoring periodic variations in magnetic properties~\cite{Krawczyk:2008jm}, such as through physical patterning like arrays of coupled nanowires~\cite{Gubbiotti:2007ea, Ding:2011gf, Huber:2013hw}, etched grooves~\cite{Chumak:2008es, Chumak:2009eg, Chumak:2009ko}, antidot lattices~\cite{Ulrichs:2010jh, Neusser:2010hr, Neusser:2011bd, Bali:2012cl, Schwarze:2012ht, Semenova:2013ef, Kumar:2013fe, Klos:2013ja, Pal:2014bb, Gubbiotti:2015gs}, dot arrays~\cite{Tacchi:2010bd, Tacchi:2011bg}, and bicomponent materials~\cite{Lin:2011cw, Lin:2012ii, Gubbiotti:2012ih, Mruczkiewicz:2013eh}. Spatially modulated magnetic states offer an alternative to physical patterning, but require instead balancing different energy terms to stabilize desired equilibrium states. Once formed, such desired states like stripe domains, spin spirals, and skyrmion lattices, not only exhibit their own individual and collective excitation spectra~\cite{Ramesh:1988sd, Ebels:2001by, Vukadinovic:2000dk, Pini:2004sw, Prestwood:2026sr, Garst:2017rm, Srivastava:2023rd, Satywali:2021msk}, but can also, when ordered, provide a self-organized periodic energy landscape for spin-wave propagation. Here, the characteristic length scales are not defined by lithography but arise from the competition between the different aforementioned magnetic interactions, which could be dynamically controlled. This makes them particularly attractive for reconfigurable magnonics, since their period, orientation, or even topology may be modified by magnetic fields~\cite{Fallarino:2019ds}, electric fields~\cite{Srivastava:2018dsk, Fillion:2022esk}, currents~\cite{Woo:2016skc, Legrand:2017rt}, strain~\cite{Rodriguez:2012tds}, or other material parameters~\cite{Soumyanarayanan:2017tsk}. Several studies have shown that periodic magnetic textures can modify spin-wave spectra, leading to folded bands ~\cite{Banerjee:2017mb, Weber:2022tm}, mode hybridization~\cite{Szulc:2022rm}, and, in some cases, band gaps ~\cite{Ma:2015smc, Dhiman:2024rm}.

Despite these encouraging results, exploiting texture-based magnonic crystals to control spin waves remains challenging because their band structures are less straightforward to understand than those of their nanopatterned counterparts. Unlike lithographically patterned magnonic crystals, where geometry fixes the lattice constant, the periodicity and internal structure of magnetic texture-based lattices are governed by the balance of magnetic energies, which can evolve with external perturbations~\cite{Li:2015rm}, or proximity to an instability ~\cite{Grassi:2022ha, Kisielewski:2023wp}. As a result, spin waves propagating across these lattices interact with a structure in which the relevant modulation may arise from the domain period, the distance between the domain walls, or any other component of the internal structure. Consequently, the emergence of band folding, hybridization, and band gaps is complex and depends on how the spin-wave modes couple to the evolving magnetic landscape.

In this article, we present a numerical investigation of spin‑wave propagation through an ordered array of stripe domains in a ferromagnetic thin film, carried out with micromagnetic simulations. Stripe patterns arise naturally when the effective magnetization, $M_{\mathrm{eff}} = M_s - 2K_u/\mu_0 M_s$, vanishes, i.e., when the perpendicular uniaxial anisotropy $K_u$ nearly compensates the demagnetizing energy $\mu_0 M_s^2/2$. This condition can also be expressed via the quality factor $Q = 2K_u/\mu_0 M_s^2$; systems with $Q \lesssim 1$ readily support stripe phases even in zero applied field. The domains are highly responsive to in‑plane magnetic fields: such a field orients the stripes along its direction and simultaneously fixes their spatial period. As a one‑dimensional magnonic crystal, these stripe arrays enable continuous tuning of the lattice spacing $a$ by adjusting the external field, thereby allowing us to study how magnon band structures evolve with $a$ without resorting to physical patterning.

The remainder of the article is organized as follows. Section~\ref{sec:model} presents the model and simulation method. Section~\ref{sec:field} discusses the role of the applied magnetic field, which serves to orient the stripe domains and determine their spatial period. The role of cubic anisotropy is discussed in Sec.~\ref{sec:cubic}. A discussion and some concluding remarks are provided in Sec.~\ref{sec:discussion}.

\section{\label{sec:model}Model and method}
Our study concerns spin‑wave propagation in a model bismuth‑doped yttrium iron garnet (BiYIG) thin film. BiYIG has gained renewed interest for magnonics because high‑quality, sub-100 nm thick films can now be readily grown with perpendicular magnetic anisotropy while preserving low Gilbert damping and a strong magnetooptical response~\cite{Soumah:2018ul, Gouere:2022ti, Bhatti:2026tm}. These characteristics faciliate a direct comparison between spin textures, such as stripe domains and bubble arrays that are readily imaged with Kerr microscopy, and their corresponding spin‑wave spectra obtained from Brillouin light scattering spectroscopy.

We model numerically the magnetization processes in the thin film with the finite difference method using the \textsc{Mumax3} code~\cite{Vansteenkiste:2014et}. The time evolution is described by the Landau-Lifshitz equation with Gilbert damping,
\begin{equation}
\frac{d \mathbf{m}}{dt} = -\gamma_0 \mathbf{m} \times \left(\mathbf{H}_\mathrm{eff} + \mathbf{h}_\mathrm{th}\right) + \alpha \mathbf{m} \times \frac{d \mathbf{m}}{dt},
\end{equation}
where $\mathbf{m}(\mathbf{r},t)$ is the unit vector representing the magnetization, $\gamma_0 = \mu_0 g \mu_B/\hbar$  the gyromagnetic constant, and $\alpha = 0.001$ the Gilbert damping constant. In the precession term on the right-hand side, $\gamma_0 = \mu_0 g \mu_B/\hbar$ is the gyromagnetic constant and the effective field, $\mathbf{H}_\mathrm{eff} = -(1/\mu_0 M_s)\delta U/\delta m$, represents the variational derivative of the micromagnetic energy $U$ with respect to the magnetization. For $U$, we account for contributions from the exchange, dipolar, and Zeeman interactions, along with uniaxial and cubic magnetocrystalline anisotropy. To compute the spin wave spectra, we include finite temperature effects through the stochastic field term $\mathbf{h}_\mathrm{th}(\mathbf{r},t)$, which has zero mean $\langle h_{\mathrm{th},i}\rangle$ and represents a Gaussian white noise with the spectral properties
\begin{equation}
\langle h_{\mathrm{th},i}(\mathbf{r}, t) h_{\mathrm{th},j}(\mathbf{r}, t') \rangle = \frac{2 \alpha k_B T}{\mu_0 V} \delta_{ij} \delta(\mathbf{r}-\mathbf{r}') \delta(t-t'),
\end{equation}
where $i,j$ are Cartesian components and $V$ is the volume of the finite-difference cell. Time integration of the Langevin problem is performed using an adaptive time-step method~\cite{Leliaert:2017ci}. The role of the stochastic field is to populate all spin wave modes. Without loss of generality, we use $T=1$~K.

\begin{figure}
	\centering\includegraphics[width=8.5cm]{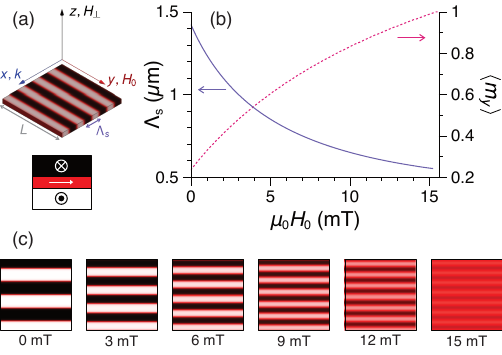}
	\caption{(a) Model geometry illustrating the simulation reference frame $(xyz)$, the applied fields $H_0$ and $H_\perp$, the simulation box length $L$, the stripe domain period $\Lambda_s$, the wave vector $k$, and the color scheme used to represent the magnetization orientation. (b) $\Lambda_s$ and mean $y$-component of the magnetization, $\langle m_y \rangle$, as a function of $H_0$. (c) Snapshots of the equilibrium magnetization state (within a subset of the simulation box, featuring $256 \times 256$ cells) for several values of $\mu_0 H_0$ with $H_\perp = 0$.}
	\label{fig:stripeperiod}
\end{figure}

The system simulated is a 22 nm‑thick BiYIG film with lateral dimensions $L\times L$, where $L \simeq 16$~$\mu$m. The volume is discretized into $1024 \times 1024 \times 1$ finite‑difference cells, with periodic boundary conditions applied along both $x$ and $y$ to emulate an infinite film. We assume an exchange constant of $A = 3.6$~pJ/m, a saturation magnetization of $M_s = 116$~kA/m, a uniaxial anisotropy constant of $K_u = 8816$~J/m$^3$, and a cubic anisotropy constant of $K_c = -464$~J/m$^3$. The easy-axis of the uniaxial anisotropy is parallel to the $z$-axis, perpendicular to the film plane, which captures the main features of BiYIG grown on a (111)-oriented substituted gadolinium gallium garnet (sGGG) substrate. An in‑plane magnetic field $\mathbf{H}_{0}=H_{0}\,\hat{\mathbf{y}}$ defines the stripe orientation, while a perpendicular bias field $\mathbf{H}_{\perp}=H_{\perp}\,\hat{\mathbf{z}}$ introduces an asymmetry between up and down domains. A schematic of this geometry is shown in Fig.~\ref{fig:stripeperiod}(a). 
%%%
%%%
The relative alignment of the cubic axes with respect to the simulation frame $(xyz)$ will be discussed in greater detail later in Sec.~\ref{sec:cubic}.

The exact value of $L$ employed in each simulation depends on the applied field $H_{0}$, because $H_{0}$ governs the stripe period. For a given $H_{0}$ (with $H_{\perp}=0$) we determine $L$ in the following way. First, we initialize the film with a stripe pattern of nominal width $L^*=16$~$\mu$m with different integer numbers of ``up'' and ``down'' domains running along $x$, separated by Bloch-type walls with magnetization along $y$. Each initial configuration is relaxed in the presence of the applied field, and the total micromagnetic energy of the relaxed state is recorded. From the resulting energy–stripe number curve we fit a quadratic function around the minimum to extract the equilibrium stripe period $\Lambda_{s}(H_{0})$. The system size is then set to $L = N_{s}\,\Lambda_{s}(H_{0})$, where $N_{s}$ is the integer that brings $L$ closest to 16~$\mu$m. For fields above the saturation field, when the ground state becomes uniformly magnetized, we simply set $L=16~\mu\text{m}$.

Figure~\ref{fig:stripeperiod}(b) displays the calculated dependence of the stripe period $\Lambda_s$ on $H_{0}$, with $H_{\perp}=0$. Over the explored range, up to the saturation field $\mu_0 H_{0,c} \approx 15.3$ mT, the period changes by almost a micron. The same panel also plots the spatially averaged magnetization component along the field direction, $\langle m_y\rangle$, illustrating the gradual reorientation of the domain moments toward $y$ as $H_{0}$ is increased. These trends are corroborated in Fig.~\ref{fig:stripeperiod}(c), which shows representative magnetization configurations for several values of $H_{0}$.

With the equilibrium stripe domain pattern established, we turn to calculating the spin‑wave spectra. We use a mode‑filtering technique that extracts the mode amplitudes directly from the simulation data with minimal post‑processing~\cite{Massouras:2024mr}. The magnetization is expanded as $\mathbf{m}_0(\mathbf{r},t)=\mathbf{m}_0(\mathbf{r})+\delta \mathbf{m}(\mathbf{r},t)$, where $\mathbf{m}_0(\mathbf{r})$ denotes the static ground state (stripe domains or uniform magnetization) and $\delta \mathbf{m}(\mathbf{r},t)$ contains the thermally driven spin‑wave fluctuations. We project these fluctuations onto a plane‑wave basis, e.g.,
\begin{equation}
\delta m_x(\mathbf{r},t)=\sum_{k}c_k(t)\,e^{ikx},
\end{equation}
with $k$ being the wave vector along $x$ and $c_k$ the associated complex mode amplitude.  Since we are interested only in the salient features of spin‑wave propagation across the stripe pattern and the ensuing magnonic crystalline behavior, we restrict our analysis to waves propagating along this direction. For a uniformly in‑plane magnetized state this corresponds to the Damon–Eshbach configuration. The filtering method yields the complex mode amplitudes $c_k(t)$ directly. To obtain the power spectrum of each mode we apply the Welch method~\cite{Welch:1967tu}. We record a time trace over 1~$\mu$s, divide it into half‑overlapping 100‑ns Hann windows, compute the discrete Fourier transform of $c_k(t)$ for each window, and then average the resulting spectra. This procedure provides an accurate estimate of the mode power as a function of frequency for every wave vector considered.

\section{\label{sec:field}Role of applied fields}
\subsection{In-plane field $H_0$}
We first focus on the role of the applied magnetic field $H_0$ (with $H_\perp = 0$), which cants the magnetization within the stripe domains toward the $y$ direction and governs the stripe period below the saturation field. We neglect for the moment the cubic anisotropy and examine only on the interplay between the exchange, dipolar, Zeeman, and uniaxial anisotropy energies.

\begin{figure}
	\centering\includegraphics[width=8.5cm]{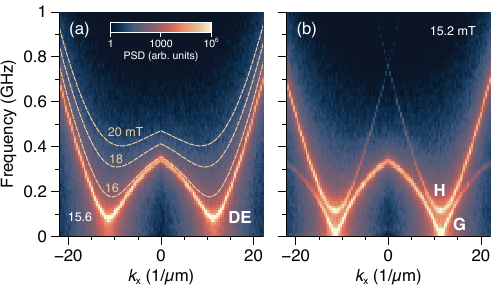}
	\caption{(a) Spin wave dispersion in the Damon-Eshbach (DE) geometry above saturation, $H_0>H_c\approx 15.3$~mT. Dashed lines represent Eq.~\ref{eq:DEmode} evaluated at three different values of $\mu_0 H_0$, while the color map shows the power spectral density obtained from simulations for $\mu_0 H_0 = 15.6$~mT. (b) Goldstone (G) and Higgs (H) modes in stripe phase at $\mu_0 H_0 = 15.2$~mT. The arrow indicates the critical wave vector $k_c = 2\pi/\Lambda_s$ at which the G mode softens.}
	\label{fig:softening}
\end{figure}

To follow how the magnonic band structure develops for spin waves propagating perpendicular to the stripes (i.e., along $x$), it is instructive to consider how the dispersion relation evolves as $H_0$ is gradually lowered through the saturation field $H_c$ from above, i.e., through the phase transition between the uniformly magnetized state and the weak stripe phase. Grassi and coworkers have shown that this transition is second–order and can be captured by a Ginzburg–Landau theory, where the softening of the Damon–Eshbach (DE) mode in the uniform state gives rise to a Goldstone mode and a Higgs‑like amplitude mode in the stripe phase~\cite{Grassi:2022ha}. This evolution is illustrated in Fig.~\ref{fig:softening}.
%%%
%%%
Just above the saturation field, the dispersion exhibits a backward-volume like character, characterized by the negative dispersion close to $k=0$, as shown in Fig.~\ref{fig:softening}. The dashed curves, representing the dispersion relation at three different values of $H_0$, correspond to the expression for dipole-exchange spin waves in the thin film limit~\cite{Kalinikos:1986tj},
\begin{equation}
\omega^2= (\omega_k + \gamma_0 N_k M_s)(\omega_k + \gamma_0 (1- N_k)M_s - \gamma_0 H_K),
\label{eq:DEmode}
\end{equation}
where $\omega_k = \gamma_0 H_0 + (2\gamma A/M_s)k^2$, $N_k = 1-(1-\exp(|k|d)/|k|d$, and $H_K = 2K_u/\mu_0 M_s$. The color map represents the wave-vector-resolved power spectral density at $\mu_0 H_0 = 15.6$~mT, from which the dispersion relation can be read directly. As $H_0$ is lowered, the minimum in this dispersion deepens. At the saturation field, the frequency at this minimum becomes zero and the critical wave vector $k_c$ at which this occurs defines the stripe period~\cite{Leaf:2006ba, Grassi:2022ha}, i.e., $k_c = 2\pi / \Lambda_s$. This soft mode corresponds to the Goldstone mode, which persists at all fields below saturation because it reflects the translational symmetry of the stripe phase along $x$. A second finite‑frequency branch appears in the stripe phase; this amplitude fluctuation corresponds to a Higgs‑like mode~\cite{Grassi:2022ha}. In what follows, we will use the labels Goldstone and Higgs to refer to these two frequency branches in $\omega(k)$.

\begin{figure*}
	\centering\includegraphics[width=17cm]{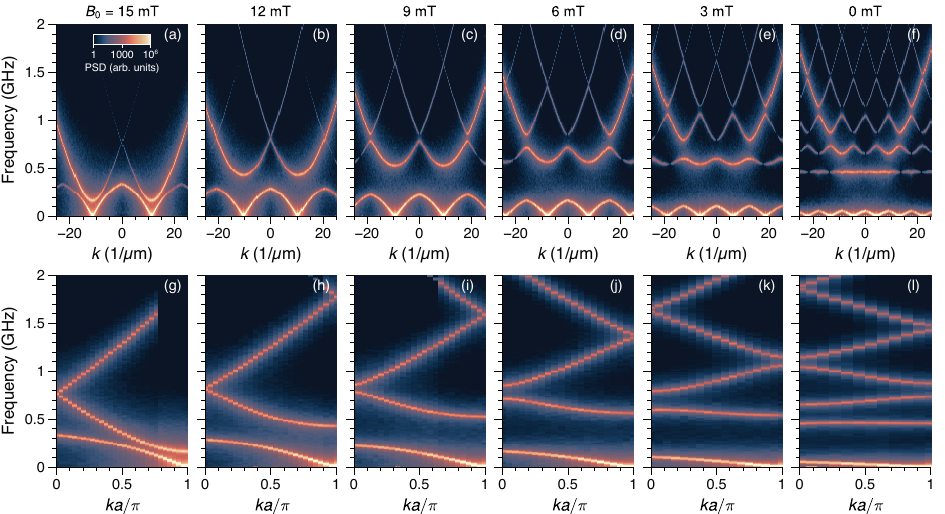}
	\caption{Color maps of the power spectral density showing the spin wave dispersion relations for different values of the applied field, $B_0 = \mu_0 H_0$. (a)-(f) Dispersion relation for spin wave propagating along the $x$ direction, for decreasing values of $B_0$. (g)-(l) Dispersion relations in (a)-(f) represented in the irreducible Brillouin zone, with $a=\Lambda_s/2$ being a field-dependent quantity [see Fig.~\ref{fig:stripeperiod}(b)].}
	\label{fig:PSDvarBx}
\end{figure*}

Figure~\ref{fig:PSDvarBx} illustrates how the spin‑wave dispersion further evolves when the applied field is decreased inside the stripe phase.
%%%
%%%
As the field is lowered toward zero in Figs.~\ref{fig:PSDvarBx}(a)-\ref{fig:PSDvarBx}(f), three clear trends appear. First, the frequency separation between the Higgs and Goldstone branches widens with decreasing field; this reflects that the domain‑wall resonance modes (G) are largely field‑independent while the domain resonances (H) shift to higher frequencies as the domains acquire a stronger perpendicular component. Second, more points along the $k$-axis at which the Goldstone branch softens emerge, directly mirroring the increase of the stripe period with decreasing field [see Fig.~\ref{fig:stripeperiod}(b)]. Third, and most significant for the magnonic crystalline nature of the stripe phase, the number of band gaps within the Higgs branch grows markedly as the field is reduced.

We can paint a clearer picture of how the band gaps emerge by recasting the dispersion relations in Figs.~\ref{fig:PSDvarBx}(a)–\ref{fig:PSDvarBx}(f) into an irreducible Brillouin‑zone (IBZ) representation that incorporates the field‑dependent periodicity.  The resulting maps are shown in Figs.~\ref{fig:PSDvarBx}(g)–\ref{fig:PSDvarBx}(l). Because the stripe domains constitute a lattice along only one spatial direction and spin‑wave propagation is reciprocal along this axis, there exist just two high‑symmetry points to consider: $k=0$, which corresponds to the $\Gamma$ point, and $k=\pi/a = 2\pi/\Lambda_{s}$, which corresponds to the $X$ point at the edge of the first Brillouin zone.  Note that the lattice period is half the domain period, $a=\Lambda_{s}/2$; this underscores the fact that the crystal structure is defined by the periodic array of domain walls rather than by the domains themselves. Within this framework, the evolution of the Higgs branch as $H_{0}$ varies becomes clearer. At high fields [Figs.~\ref{fig:PSDvarBx}(g) and \ref{fig:PSDvarBx}(h)] no band gaps appear in the Higgs branch; we merely observe the familiar folding of a freely propagating mode in the reduced zone.  As $H_{0}$ is progressively lowered, a gap opens at the Brillouin‑zone center ($k=0$), and its width grows with decreasing field [Figs.~\ref{fig:PSDvarBx}(i)–\ref{fig:PSDvarBx}(j)].  With further reduction of the field, additional gaps also appear at the zone edge ($k=\pi/a$).  At zero applied field the lowest Higgs branch collapses into a nearly dispersionless mode, while sizable band gaps separate higher‑frequency branches.

\begin{figure}
	\centering\includegraphics[width=6.5cm]{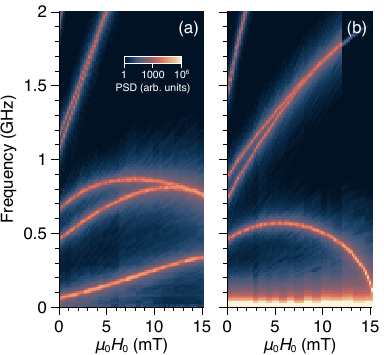}
	\caption{Color maps of the power spectral density at the (a) $k=0$ and (b) $k=\pi/a$ points as a function of applied magnetic field $H_0$.}
	\label{fig:fgx}
\end{figure}

To follow the evolution of the band gaps with applied field we plot the power spectral density (PSD) at the two symmetry points, $k=0$ and $k=\pi/a$, for different values of $\mu_0 H_0$ as a color map.  This is shown in Fig.~\ref{fig:fgx}.  
%%%
%%%
At $k=0$ the lowest‑frequency branch is the Goldstone branch, while higher branches correspond to the folded Higgs band.  A gap opens in the first Higgs branch around 11~mT as the field is reduced from saturation; this gap widens progressively as the field is reduced to zero.  The second Higgs branch likewise develops a gap, but at a different field value.  These features should be directly observable in broadband ferromagnetic‑resonance experiments. At the Brillouin‑zone edge ($k=\pi/a$) the Goldstone mode remains at zero frequency and several Higgs branches appear above it.  Interestingly, the first Higgs branch stays gapless over the entire field range, whereas a gap opens in the second branch once $\mu_{0}H_{0}$ falls below approximately 7~mT. This figure highlights how the band structure of the Higgs‑branch can be tailored by varying the in-plane field.

\subsection{Perpendicular field $H_\perp$}
Let us now examine the influence of a perpendicular field $H_{\perp}$ on the magnonic band structure.  For $H_{\perp}>0$, it is energetically favorable for the ``up'' domain to expand while the ``down'' domain contracts, thereby minimizing the extra Zeeman contribution. We denote by $\Lambda_{d}^{+}$ the width of an ``up'' domain, i.e., the portion along $x$  within one stripe period in which $m_{z}>0$, and analogously define $\Lambda_{d}^{-}$ for a ``down'' domain ($m_{z}<0$).  By construction, $\Lambda_{d}^{+}+\Lambda_{d}^{-}= \Lambda_{s}$. Although $\Lambda_{s}$ acquires a weak dependence on $H_{\perp}$, our analysis of the band structure focuses on the evolution within the IBZ, so any change in the domain period is accounted for implicitly.

\begin{figure}
	\centering\includegraphics[width=8.5cm]{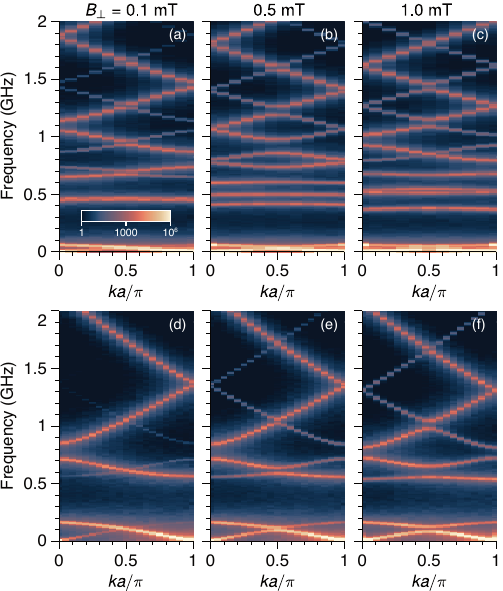}
	\caption{Color maps of the power spectral density in the reduced Brillouin zone assuming a lattice period of $a = \Lambda_s/2$ under an in-plane field of (a-c) $\mu_0 H_0 = 0$ and (d-f) $\mu_0 H_0 = 6$~mT. The emergence of reflected bands about and band gaps at $k=\pi/2a$ with $\mu_0 H_\perp > 0$ indicates a doubling of the effective lattice constant, i.e., $a' = \Lambda_s$.}
	\label{fig:PSDvarBz}
\end{figure}

Figure~\ref{fig:PSDvarBz} displays the calculated dispersion for several values of $H_{\perp}$ and a fixed in‑plane field $H_{0}$.  
%%%
%%%
The presence of a finite perpendicular field ($H_{\perp}>0$) gives rise to reflected bands centered at $k=\pi/2a$, whose intensity grows with increasing $H_{\perp}$.  This is evident for the two cases shown, namely for $\mu_0 H_0 = 0$~mT in Figs.~\ref{fig:PSDvarBz}(a)-\ref{fig:PSDvarBz}(c) and 6~mT in Figs.~\ref{fig:PSDvarBz}(d)-\ref{fig:PSDvarBz}(f), respectively.  Moreover, gaps at $k=\pi/2a$ open progressively as \(H_{\perp}\) is raised.  These features are consistent with an effective doubling of the lattice constant, $a' = \Lambda_{s}$, which occurs because $\Lambda_{d}^{+} \neq \Lambda_{d}^{-}$ whenever $H_{\perp}\neq 0$.

\begin{figure}
	\centering\includegraphics[width=8.5cm]{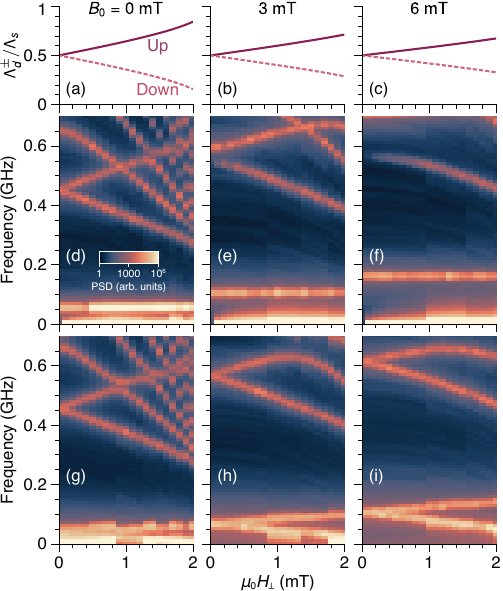}
	\caption{(a-c) Variation in relative size of ``up'' ($\Lambda^+_d$) and ``down'' ($\Lambda^-_d$) domains, with respect to the domain period $\Lambda_s$, as a function of perpendicular field $H_\perp$. (d-f) Color maps of the power spectral density at the $k=0$ point as a function of $H_\perp$. (g-i) Color maps of the power spectral density at the $k=\pi/2a=\pi/\Lambda_s$ point as a function of $H_\perp$ (\emph{cf} Fig.~\ref{fig:PSDvarBz}). All quantities are shown for three values of $B_0 = \mu_0 H_0$: (a,d,g) 0 mT, (b,e,h) 3 mT, and (c,f,i) 6 mT.}
	\label{fig:fgxbz}
\end{figure}

Figure~\ref{fig:fgxbz} shows how the asymmetry between $\Lambda_d^{+}$ and $\Lambda_d^{-}$ correlates with band gaps at the Brillouin‑zone center $k=0$ and at the newly emergent symmetry point $k=\pi/2a=\pi/a'=\pi/\Lambda_s$.  
%%% 
%%%
The relative sizes of the ``up'' and ``down'' domains, $\Lambda_d^{\pm}/\Lambda_s$, are plotted as a function of $H_\perp$ for three values of the in‑plane field $\mu_0 H_0$ in Figs.~\ref{fig:fgxbz}(a)–\ref{fig:fgxbz}(c). All cases exhibit a similar trend, with the largest asymmetry between $\Lambda_d^{+}$ and $\Lambda_d^{-}$ occurring for $H_{0}=0$. The corresponding changes in the PSD at $k=0$ are shown in Figs.~\ref{fig:fgxbz}(d)–\ref{fig:fgxbz}(f), with emphasis on the low‑frequency bands.  Because of the mirroring at $k=\pi/2a$, the Goldstone mode that originally lies at $k=\pi/a$ is mapped onto $k=0$, producing two nearly horizontal lines below 200~MHz in these plots.  A band gap opens in the first‑order Higgs branch, and additional band crossings appear as $H_{\perp}$ increases. At the new symmetry point $k=\pi/2a=\pi/\Lambda_s$, gaps open for both the Goldstone and Higgs branches, as illustrated in Figs.~\ref{fig:fgxbz}(g)–\ref{fig:fgxbz}(i).  A similar interplay with higher‑order descending branches is observed at $H_0 = 0$ [Fig.~\ref{fig:fgxbz}(g)].

\section{\label{sec:cubic}Role of cubic anisotropy}
\begin{figure}
	\centering\includegraphics[width=6cm]{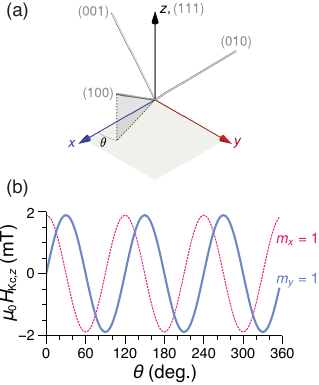}
	\caption{(a) Geometry of the cubic anisotropy axes with respect to the film plane and simulation reference frame. (b) Perpendicular component of the effective field, $H_{\mathrm{Kc},z}$, as a function of the orientation of the cubic axes for a uniform magnetization state oriented along $x$ and $y$, respectively.}
	\label{fig:KcCoords}
\end{figure}

Having shown that the Higgs‑branch band gaps can be tuned by varying applied magnetic fields, we now discuss how cubic anisotropy can achieve a similar effect.  With a fixed value of $K_c$ we vary only the orientation of its principal axes relative to the in‑plane field (and hence to the stripe direction). Figure~\ref{fig:KcCoords}(a) defines our convention: the angle $\theta$ is measured between the $x$-axis and one of the cubic axes, e.g. [100], projected onto the film plane. 
%%%
%%%
A useful way to anticipate the effect of cubic anisotropy on the stripe texture is to examine how the effective cubic anisotropy field varies with rotations of the $[100]$ axis. For magnetization states with $m_z = \pm 1$, the out‑of‑plane component of this field, $\mu_{0}H_{\mathrm{Kc},z}$, is independent of $\theta$ and remains approximately $\pm 5.3$~mT; it merely renormalizes the effective perpendicular anisotropy.  When $m_{y}=1$, however, $\mu_{0}H_{\mathrm{Kc},z}$ displays a $\sin(3\theta)$ dependence with an amplitude of about 2~mT and a period of $120^{\circ}$, reflecting the three‑fold symmetry of the cubic term [see Fig.~\ref{fig:KcCoords}(b)].  For completeness we also show that for $m_{x}=1$ the dependence follows $\cos(3\theta)$. Because the equilibrium stripe pattern is mainly governed by the $y$- and $z$-components of the magnetization, the relation between $\theta$ and $H_{\mathrm{Kc},z}$ implies that the cubic anisotropy should not alter the spin‑wave dispersion at the high‑symmetry angles $\theta=n\pi/6$ (with integer $n$).  Drawing on what we learned about how a perpendicular field $H_{\perp}$ influences the band structure, we can expect that cubic anisotropy will similarly introduce an asymmetry between the ``up'' and ``down'' domain widths, as discussed in Ref.~\onlinecite{Prestwood:2026sr}.

\begin{figure}
	\centering\includegraphics[width=8.5cm]{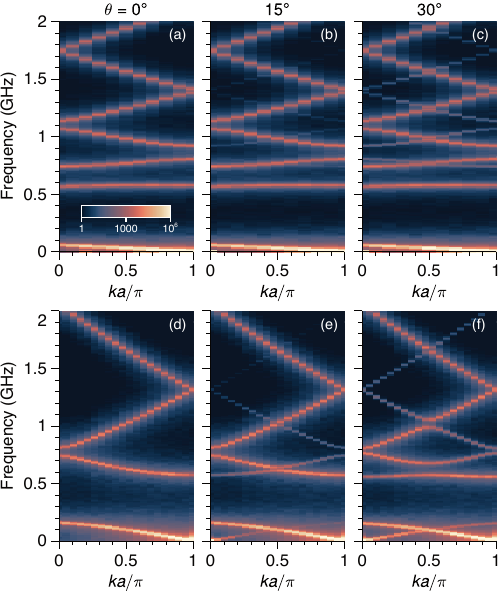}
	\caption{Color maps of the power spectral density in the IBZ assuming a lattice period of $a = \Lambda_s/2$ under an in-plane field of (a-c) $\mu_0 H_0 = 0$ and (d-f) $\mu_0 H_0 = 6$~mT, for three different values o the cubic axis orientations $\theta$: (a,d) 0$^\circ$, (b,e) 15$^\circ$, and (c,f) 30$^\circ$. The emergence of reflected bands about and band gaps at $k=\pi/2a$ with $0 < \theta < 60^\circ$ indicates a doubling of the effective lattice constant, $a'=\Lambda_s$, analogous to the case for $H_\perp \neq 0$ (Fig.~\ref{fig:PSDvarBz}).}
	\label{fig:PSDvarKc}
\end{figure}

Let us now examine how the magnon spectrum evolves as a function of cubic anisotropy axes. As before we restrict our discussion to the IBZ so that we capture the most salient features.  Figure~\ref{fig:PSDvarKc} shows the band structure for several orientations $\theta$ of the $[100]$ axis with respect to $x$, for two values of $B_{0}$.
%%%
%%%
First consider the case of zero applied field [Figs.~\ref{fig:PSDvarKc}(a)–\ref{fig:PSDvarKc}(c)].  In this situation only the spins inside the domain walls possess a component along $y$; consequently the cubic anisotropy is expected to only affect the moments within these regions. For $\theta=0^\circ$ [Fig.~\ref{fig:PSDvarKc}(a)] the band structure essentially reproduces that of a purely uniaxial film at zero field [Fig.~\ref{fig:PSDvarBx}(l)], as anticipated from our analysis of the effective field. When $\theta$ is increased to $15^\circ$ and $30^\circ$ [Figs.~\ref{fig:PSDvarKc}(b), \ref{fig:PSDvarKc}(c)], a second set of bands gradually appears. These are weaker replicas of the original modes, reflected about $k=\pi/2a$.  They become most pronounced at higher frequencies in Fig.~\ref{fig:PSDvarKc}(c), where $\mu_{0}H_{\mathrm{Kc},z}$ is strongest for large values of $m_y$. Further increasing the angle causes the secondary bands to fade and disappear entirely by $\theta=60^\circ$.  A full calculation over the entire range $0\leq\theta<2\pi$ confirms that this behavior repeats every $\pi/3$, indicating that it depends only on the magnitude $|H_{\mathrm{Kc},z}|$, not its sign.

With a larger in‑plane field ($\mu_{0}H_0 = 6~\text{mT}$) the effect becomes markedly stronger, as displayed in Figs.~\ref{fig:PSDvarKc}(d)–\ref{fig:PSDvarKc}(f). The mirrored secondary bands reappear, just as they do when $H_\perp \neq 0$. These reflected branches intersect the original spectrum at $k=\pi/2a = \pi/\Lambda_s$, giving rise to new band gaps.  The most pronounced gap occurs in the lowest Higgs branch, as seen in Fig.~\ref{fig:PSDvarKc}(f), where the cubic anisotropy exerts its greatest influence. This accompanied by a reduction in the gap at $k=0$.

\begin{figure}
	\centering\includegraphics[width=8.5cm]{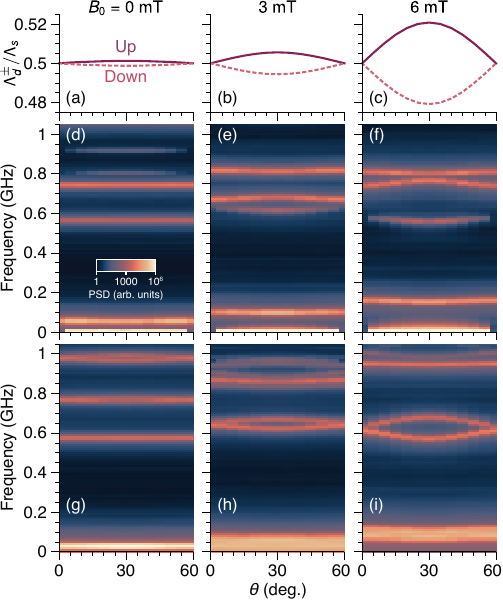}
	\caption{(a-c) Variation in relative size of ``up'' ($\Lambda^+_d$) and ``down'' ($\Lambda^-_d$) domains, with respect to the domain period $\Lambda_s$, as a function of cubic axis orientation $\theta$. (d-f) Color maps of the power spectral density at the $k=0$ point. (g-i) Color maps of the power spectral density at the $k=\pi/2a=\pi/\Lambda_s$ point (\emph{cf} Fig.~\ref{fig:PSDvarKc}). All quantities are shown for three values of $\mu_0 H_0$: (a,d,g) 0, (b,e,h) 3~mT, and (c,f,i) 6~mT.}
	\label{fig:fgx_varKc}
\end{figure}

We now turn our attention to how the induced gaps vary with orientation of the cubic axes and field strength, with a comparison with changes in the relative domain sizes. Figures~\ref{fig:fgx_varKc}(a)-\ref{fig:fgx_varKc}(c) display the relative change in $\Lambda_d^{\pm}$ as a function of $\theta$ for three different values of $H_0$.
%%%
%%%
For all cases considered, the ``up'' and ``down'' domains have the same size at $\theta = 0^\circ$ and $60^\circ$, with the largest asymmetry occurring at $\theta = 30^\circ$. This is consistent with our argument above based on the angular dependence of $H_{\mathrm{Kc},z}$.

Figures~\ref{fig:fgx_varKc}(d)-\ref{fig:fgx_varKc}(f) show how the PSD at $k=0$ changes as the cubic‑anisotropy axis is rotated for three values of the in‑plane field $H_0$. In zero applied field the bands are almost unchanged, except for the mapping of the Goldstone mode onto $k=0$, as discussed earlier for $H_\perp = 0$.  When a finite $H_{0}$ tilts the magnetization into the film plane, noticeable modifications appear.  The gap between the original and mirrored branches of the first Higgs band is largest at $\theta =30^{\circ}$, while the separation between the two original Higgs bands is smallest at the same angle. At the new symmetry point $k=\pi/2a=\pi/\Lambda_s$ [Figs.~\ref{fig:fgx_varKc}(g)–\ref{fig:fgx_varKc}(i)] the gaps disappear for $\theta=0^{\circ}$ and $60^{\circ}$, and attain a maximum at $\theta=30^{\circ}$; this mirrors the angular dependence of $H_{\mathrm{Kc},z}$ shown in Fig.~\ref{fig:KcCoords}(b).  The same trend is reflected in the variation of the ``up''‑ and ``down''‑domain widths, $\Lambda_d^{+}$ and $\Lambda_d^{-}$, plotted in Figs.~\ref{fig:fgx_varKc}(a)–\ref{fig:fgx_varKc}(c).

Similarly, for angles $60^{\circ}<\theta<120^{\circ}$ the sign of $H_{Kc,z}$ reverses, causing the ``down'' domain to grow at the expense of the ``up'' domain.  However, because of inversion symmetry, the resulting band‑structure modifications are identical to those seen for $0^{\circ}\leq\theta\leq60^{\circ}$.  In general, the influence of the cubic‑anisotropy axis is greatest when $H_{0}$ is large, since a stronger in‑plane field tilts the magnetization more toward the $y$ direction and thereby amplifies the contribution from $H_{\mathrm{Kc},z}$.

\section{\label{sec:discussion}Discussion and concluding remarks}

Like all waves propagating in a periodic potential, the non‑trivial band structures that emerge for spin waves traversing stripe textures (Figs.~\ref{fig:PSDvarBx} and \ref{fig:PSDvarKc}) arise from Bragg reflections due to scattering off the regular array of domain walls. Band gaps survive even as the field pulls the domain magnetization toward the film plane, demonstrating that the phenomenon is robust for both strong and weak stripe textures.

The gaps are most pronounced in the absence of an applied field [Figs.~\ref{fig:PSDvarBx}(f) and \ref{fig:PSDvarBx}(l)] when cubic anisotropy is neglected.  In this situation the film hosts a perfectly periodic lattice of $180^{\circ}$ Bloch domain walls [Fig.~\ref{fig:stripeperiod}(c)].  The seminal work of Winter~\cite{Winter:1961hw} established that Bloch walls represent reflectionless potentials for spin waves, being a special case of the P{\"o}schl–Teller potential in one‑dimensional wave propagation~\cite{Poeschl:1933bz, Infeld:1951ft}.  Reflectionless interfaces should therefore transmit waves without Bragg reflection by defintion, leaving the band structure unchanged from that of a free particle.  This conjecture has been confirmed numerically in systems containing only exchange interactions and uniaxial and transverse anisotropies~\cite{Borys:2015ba}, with the latter representing a local form of the demagnetizing field. In this light, the gaps seen in Figs.~\ref{fig:PSDvarBx}(f) and \ref{fig:PSDvarBx}(l) may appear counterintuitive and contradictory.  The resolution is that the reflectionless property holds only when exchange and local anisotropies are the sole energetic terms; the inclusion of anisotropic, nonlocal terms such as dipolar or Dzyaloshinskii–Moriya interactions (DMI) introduces finite scattering between Winter magnons and the domain‑wall potential.  The same numerical work has indeed shown band gaps in arrays of N{\'e}el-type domain walls with DMI~\cite{Borys:2015ba}, while DMI-like unidirectional domain‑wall magnons can also appear when dipolar interactions are accounted for in Bloch walls~\cite{Henry:2019dp}.  These results imply that non‑trivial band structures are inevitable for ferromagnetic stripe textures, since dipolar interactions stabilize these configurations in the first place.

Another striking feature is the appearance of reflected bands and gaps at $k=\pi/2a=\pi/\Lambda_s$, as exemplified in Figs.~\ref{fig:PSDvarBz}(f) and \ref{fig:PSDvarKc}(f).  The strength of these gaps correlates with the asymmetry between ``up'' and ``down'' domains, which can be viewed as a Peierls‑type distortion of the domain wall lattice~\cite{Peierls:1955}.  In the classic picture of this phenomenon in a one‑dimensional crystal, an ion displaces toward one neighbor while moving away from the other, creating alternating short and long bonds and doubling the lattice constant.  For a conductor this coupling between electrons and the lattice opens gaps at $k=\pm\pi/2a$.  The same mechanism operates here: either $H_\perp$ or $\mu_0 H_{\mathrm{Kc},z}$ breaks the symmetry between ``up'' and ``down'' domains, so one domain grows while the other shrinks.  The spacing between consecutive domain walls then alternates, just as bond lengths alternate in a dimerized lattice.  In this light, spin waves traversing stripe textures therefore not only provide a laboratory for Higgs and Goldstone modes in second‑order phase transitions~\cite{Grassi:2022ha}, but may also offer an intriguing platform to explore Peierls‑driven metal–insulator physics in one‑dimensional crystals.

\begin{acknowledgments}
We thank Thibaut Devolder, Ping Che, and Pawe{\l} Gruszecki for fruitful discussions. This work was partially supported by the Agence Nationale de la Recherche (France) under contract numbers  ANR-22-CE30-0014 (DeMIuRGe) and ANR-24-CE24-0354 (ANTIPASTI). VL acknowledges financial support from the EOBE Doctoral School of Universit{\'e} Paris-Saclay.
\end{acknowledgments}

\bibliography{articles}

\end{document}